\documentclass[report,epsfig,pre]{revtex4}

\usepackage{graphicx}
\usepackage{amssymb}
\usepackage{amsmath}
\usepackage{color}

\newcommand{\figsizeone}{0.48}

\begin{document}

\draft
\title{Reconfiguration of quantum states in $\mathcal PT$-symmetric quasi-one dimensional lattices}
\author{Jung-Wan Ryu, Nojoon Myoung, and Hee Chul Park}
%\email{}
\affiliation{Center for Theoretical Physics of Complex Systems, Institute for Basic Science, Daejeon 34051, South Korea}
%\date{\today}

\begin{abstract}
We demonstrate mesoscopic transport through quantum states in quasi-1D lattices maintaining the combination of parity and time-reversal symmetries by controlling energy gain and loss. We investigate the phase diagram of the non-Hermitian system where transitions take place between unbroken and broken $\mathcal{PT}$-symmetric phases via exceptional points. Quantum transport in the lattice is measured only in the unbroken phases in the energy band--but not in the broken phases. The broken phase allows for spontaneous symmetry-broken states where the cross-stitch lattice is separated into two identical single lattices corresponding to conditionally degenerate eigenstates. These degeneracies show a lift-up in the complex energy plane, caused by the non-Hermiticity with $\mathcal{PT}$-symmetry.
\end{abstract}

\maketitle

~~~~~ Non-Hermiticity has attracted great interest, both theoretical and experimental, in open systems with energy gain and loss \cite{Moi11}. While we consider a physical object to be isolated, connections with the environment strongly influence the physical properties of quantum and classical systems by breaking both energy and flux conservation. From a mathematical viewpoint, these open systems can be described by non-Hermitian formalism, with closed ones by Hermitian formalism.
After non-Hermiticity was found to arise in vortex dynamics \cite{Nel93}, Hatano and Nelson provided a prototypical model of non-Hermitian systems that considers the hopping dynamics of a particle on an Anderson lattice in the presence of a constant imaginary vector potential \cite{Hat96}. Since then, research has continually attempted to integrate this model into quantum systems \cite{Bro97, Sil98, Hat98}.

One of the interesting properties of non-Hermiticity is the existence of a non-Hermitian degeneracy, called an exceptional point (EP), where two complex eigenvalues and corresponding eigenstates coalesce \cite{Kat96, Hei12}. The EPs allow for interesting phenomena such as level crossing, geometric phases, and chirality without the symmetry typically necessary for accidental degeneracy. The EPs and their related phenomena have been studied in open systems such as atomic spectra in fields \cite{Lat95, Car07}, microwave cavity experiments \cite{Dem01, Dem03}, and chaotic optical microcavities \cite{Lee08, Lee09}, among others. In addition, EPs have also been observed in purely classical systems such as coupled oscillators with damping \cite{Hei04, Ste04, Ryu15}. 

Among non-Hermitian systems, parity-time ($\mathcal{PT}$)-symmetric systems exhibit a transition from an unbroken $\mathcal{PT}$- symmetric phase to a broken phase via EPs \cite{Ben98, Ben07}. $\mathcal{PT}$-symmetry is protected in non-Hermitian systems with a balance of energy gain and loss represented by the commutation relation $[\mathcal{H},\mathcal{PT}] = 0$, where $\mathcal{H}$ is a Hamiltonian. While the unbroken phase contains normal dispersive states with real eigenenergies, the broken phase allows for spontaneous symmetry breaking in the eigenstates with complex energies. Many $\mathcal{PT}$-symmetric systems have been explored in several fields, including optical cavities \cite{El07, Guo09, Rue10}, electronic circuits \cite{Sch11}, atomic physics \cite{Jog10}, magnetic metamaterials \cite{Laz13}, and photonic lattices \cite{Mak08, Reg12, Cer16}. 
Despite such a range, there have been few studies concerning quantum transport in $\mathcal{PT}$-symmetric systems.

In this work, we demonstrate that a bandgap is realized in $\mathcal{PT}$-symmetric quasi-1D lattices by quantum measurement through external leads. The bandgap is shown to be controlled by external parameters such as energy gain and loss, but not by inherent properties such as geometric shape or defects. Moreover, the spontaneously symmetry-broken states inside the bandgap are measured by controlling non-Hermitian parameter breaking $\mathcal{PT}$-symmetry. Finally, we devise a mechanism for reconfiguration of quantum states that is related to spontaneous symmetry breaking through non-Hermiticity.

\begin{figure}
\centering
\includegraphics[width=0.7\columnwidth]{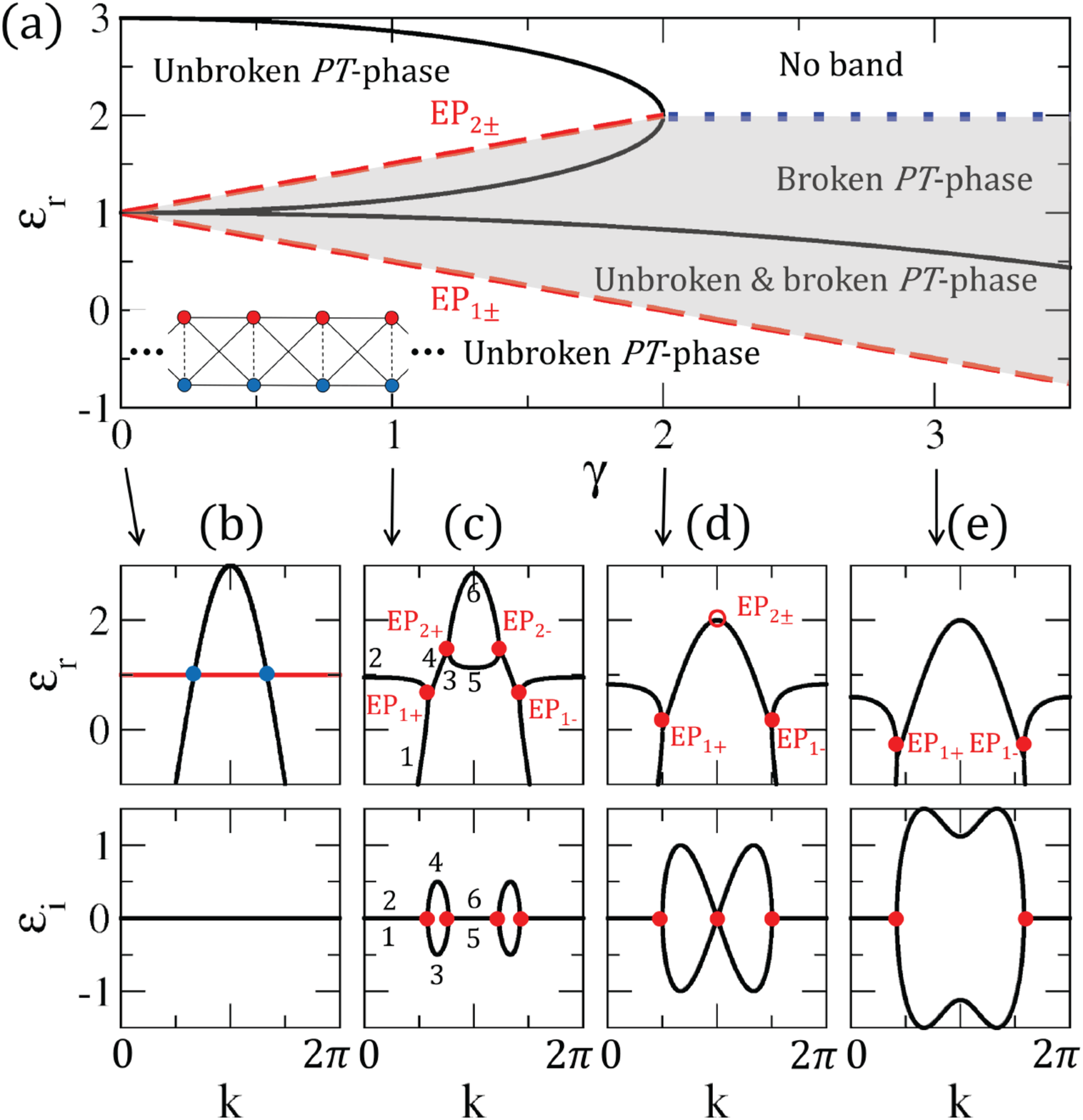}
\caption{(a) Phase diagram of $\mathcal{PT}$-symmetric CS lattice. Black curves represent the band edges of unbroken $\mathcal{PT}$-symmetric phase $\varepsilon^*$. There are two red dashed lines of EPs inside the unbroken $\mathcal{PT}$-symmetric phase regions: $EP_1\pm$ and $EP_2\pm$. The shaded region indicates the broken $\mathcal{PT}$-symmetric phase, which includes a coexistent region with unbroken and broken phases. The blue dotted line is the critical energy $\varepsilon_{c}=2d$. (inset) $\mathcal{PT}$-symmetric CS lattice with coupling strengths $t$ (dashed lines) and $d$ (solid lines). The complex energies on the upper sites (red dots) and lower sites (blue dots) correspond to energy gain $+i \gamma /2$ and energy loss $-i \gamma /2$, respectively. (b) Complex energy band for the CS lattice. The constant energy band assigned by the solid red line is the flat band, and the blue dots represent accidental degenerate points. (c)-(e) Complex energy bands with different non-Hermitian imbalanced potentials. The red dots are exceptional points. We fix $t=d=1$ throughout this paper.
}
\label{fig1}
\end{figure}

%\section{Bands of $\mathcal{PT}$-symmetric cross-stitch lattices}
~~~~~ {\it Bands of $\mathcal{PT}$-symmetric cross-stitch lattices} - 
When we consider a cross-stitch (CS) lattice, as shown in the inset of Fig.~\ref{fig1}, the non-Hermitian Hamiltonian can be expressed on the up/down basis of Pauli matrix $\pmb{\sigma}=(\sigma_x,\sigma_z)$ and identity matrix $\sigma_0$, and in terms of vector field $\pmb{h}(k)=(-t-2d\cos{k},\delta/2+i\gamma/2)$ and extra term $h_0=-2d\cos{k}$, as
\begin{eqnarray}
H(k)=\pmb{h}(k)\cdot\pmb{\sigma}+h_0(k)\sigma_0,
\end{eqnarray}
where, using Bloch theorem, $t$ and $d$ are hopping constants of intra and inter unit cells, and $\delta$ and $\gamma$ are real asymmetric energy and balanced gain and loss on site, respectively. The eigenvalues of the Hamiltonian are
\begin{eqnarray}
\nonumber
\varepsilon_{\pm} & = & \pm|\pmb{h}(k)|+h_0(k) \\
& = & -2d \cos{k} \pm \sqrt{(t+2d \cos{k})^2 + \left(\frac{\delta+ i \gamma}{2}\right)^2},
\label{energy}
\end{eqnarray}
which are complex, as $\varepsilon = \varepsilon_r + i \varepsilon_i$.

In the case of a symmetric lattice, where $\delta=\gamma=0$, the lattice can be considered as a combination of two identical 1D chains on which hopping strength is $d$ so that the eigenenergy of a single chain is $-2d \cos{k}$. 
The eigenenergies of symmetric lattices are $t$ and $-t - 4d \cos{k}$, which are flat and dispersive bands, as illustrated in Fig.~\ref{fig1} (b), due to the antisymmetric and symmetric configuration between the two chains, respectively. 
As $t$ increases, flat and dispersive bands lift and lower, respectively. As $d$ increases, the dispersive band widens.
The two bands are completely decoupled, which results in degenerate points because the system protects even/odd parities for reflection symmetry through a horizontal axis. The odd symmetry of the flat band generates compact localized states with nonzero amplitude only at a finite number of unit cells due to destructive interference, while even parity generates propagated states \cite{Fla14, Ge15, Mai16}.

Following symmetry breaking, the two states are mixed, and thus the band structure is modified into dispersive bands with a finite energy gap and no flat band. There are three ways to break the reflection symmetry of on-sites through the combination of real and imaginary imbalanced potentials based on the Hamiltonian above. For the potentials, we neglect both real and imaginary overall on-site potentials as they merely play the role of band shifts through the real and imaginary directions, respectively.
In the case of Hermitian imbalanced potential, where $\delta \neq 0$ and $\gamma=0$, the broken reflection symmetry causes dispersive bands with bandgaps due to the Hermitian degeneracy lifting we expect (see Supplementary Information I).
For $\mathcal{PT}$-symmetric imbalanced potential, where $\delta = 0$ and $\gamma \neq 0$, the Hermitian degenerate point splits into two non-Hermitian degenerate points, or EPs, distinguishing the real parts of the eigenenergies from the Hermitian cases. The corresponding band structures are shown in Fig.~\ref{fig1} (b)–-(e). There are transitions between unbroken and broken phases at the EPs in which two eigenenergies degenerate into exactly the same states. The broken phases have two complex conjugate eigenenergies, but the unbroken phases have two real eigenenergies in spite of the non-Hermitian Hamiltonian. The real parts of the eigenenergies in the broken phase range always degenerate into eigenenergies of the 1D chain, and these eigenstates are localized only on the upper or lower chain as a consequence of spontaneous symmetry breaking.
Lastly, general non-Hermitian imbalanced potential, where $\gamma \neq 0$ and $\delta \neq 0$, generates complex eigenenergies which have no EPs in the band structures because the overall real imbalanced on-site potential breaks both non-Hermitian and Hermitian degeneracies.

%Fig1%

Let us concentrate on $\mathcal{PT}$-symmetric quasi-1D lattices including EPs. 
In Fig.~\ref{fig1} (c), as aforementioned, two real energies (1 and 2) change into two complex conjugated energies (3 and 4) via $EP_{1+}$, and then back into two real energies (5 and 6) again via $EP_{2+}$ with tracing energies following the momentum. The eigenstates corresponding to bands 1 and 2 are in the unbroken phase where the wave functions occupy both up and down states; the eigenstates with bands 3 and 4, however, are in the broken phase where the wave functions only occupy up or down states.

Figure~\ref{fig1} (a) is a phase diagram for the $\mathcal{PT}$-symmetric CS lattice in ($\gamma$, $\varepsilon_r$) space, where the edges of the bands are separated.
The unbroken bands are constructed by the upper band edge at $k=\pi$ and the lower band edge at $k=0$, as follows,
\begin{equation}
\varepsilon_{ub/lb}^*(\gamma)=
   \begin{cases}
        ~~2d-\sqrt{(t-2d)^2-(\gamma/2)^2},\\
        -2d+\sqrt{(t+2d)^2-(\gamma/2)^2}.
   \end{cases}
\end{equation}
Following the Hermitian degenerate point, the non-Hermitian degenerate points are estranged from each other as
\begin{equation}
\begin{cases} \varepsilon_{EP1\pm}=t-\gamma/2,\\
\varepsilon_{EP2\pm}=t+\gamma/2,
\end{cases}
\end{equation}
where $\pm$ indicates opposite momentum $\pm k$, respectively. 
It is noted that there are overlap regions of unbroken and broken $\mathcal{PT}$-symmetric phases between EP lines and the unbroken band edges. As shown in Fig.~\ref{fig1} (d), the two EPs with opposite momentum, $EP_{2+}$ and $EP_{2-}$, merge at the Brillouin zone boundary at the critical energy gain and loss $\gamma = \gamma_{c}=2 \left|t-2d \right|$. When $\gamma >\gamma_{c}$, the evaporation of the upper band coincides with the disappearance of a pair of EPs, and a no-band region and broken phase region appear above and below the critical energy $\varepsilon_{c}=2d$, respectively, as seen in Fig.~\ref{fig1} (e).

%\section{Transport in $\mathcal{PT}$-symmetric cross-stitch lattices}

{\it Transport in $\mathcal{PT}$-symmetric coss-stitch lattices} - 
Let us consider charge transport in a $\mathcal{PT}$-symmetric quasi-1D lattice, specifically here in a CS lattice with $N$ unit cells between two leads connected to both lattice ends. The Hamiltonian for this model is given by
\begin{equation}
 H = H_{cs} + H_{lead} + H_{coupling},
\label{fb_trans}
\end{equation}
where $H_{cs}$, $H_{lead}$, and $H_{coupling}$ describe the CS lattice, leads, and coupling between the lattice and leads, respectively. Each Hamiltonian is written as follows:
\begin{eqnarray}
 H_{cs} &=& \sum_{i=1}^{N}H_0 c_{i}^{\dagger} c_{i} + \sum_{i=1}^{N-1} (H_{1} c_{i+1}^{\dagger} c_{i} + h.c.) \\
 H_{lead} &=& -\frac{V_0}{2} \sum_{j \neq 0} (b_{j+1}^{\dagger} b_{j} + h.c.) \\
 H_{coupling} &=& -g ( c_{1}^{\dagger} b_{-1} + b_{1}^{\dagger} c_{N} + h.c.),
\end{eqnarray}
where $H_0 = -t \sigma_x + (\delta/2 + i \gamma/2) \sigma_z$, $H_1 = - d (\sigma_x + \sigma_0)$, and $c_{j}^{\dagger}$ ($c_{j}$) and $b_{j}^{\dagger}$ ($b_{j}$) are creation (annihilation) operators for the lattice and leads, respectively. $V_0 / 2$ is hopping strength in the leads and $g$ is coupling strength between the ends of the CS lattice and leads.
If the wavenumber in the leads is $q$, then the wavefunction in the leads is given by
\begin{eqnarray}
 \phi_{j} =& e^{i q j} + r_0 e^{-i q j} & (j < 0) \\
          =& t_0 e^{i q j} & (j > 0),
\end{eqnarray}
where $\left|r_0\right|^2$ and $\left|t_0\right|^2$ are reflection and transmission probabilities, respectively, with $\left|r_0\right|^2 + \left|t_0\right|^2 = 1$ in $\mathcal{PT}$-symmetric cases.
Using $e^{\pm i q} = - E/V_{0} \pm i \sqrt{1 - \left|E/V_{0}\right|^2}$, we obtain transmission probabilities from the resulting $(2N+2) \times (2N+2)$ matrix, where $E$ is the incident energy from a lead \cite{hee} (see Supplementary Information II).

%Fig2%

\begin{figure}
\centering
\includegraphics[width=0.7\columnwidth]{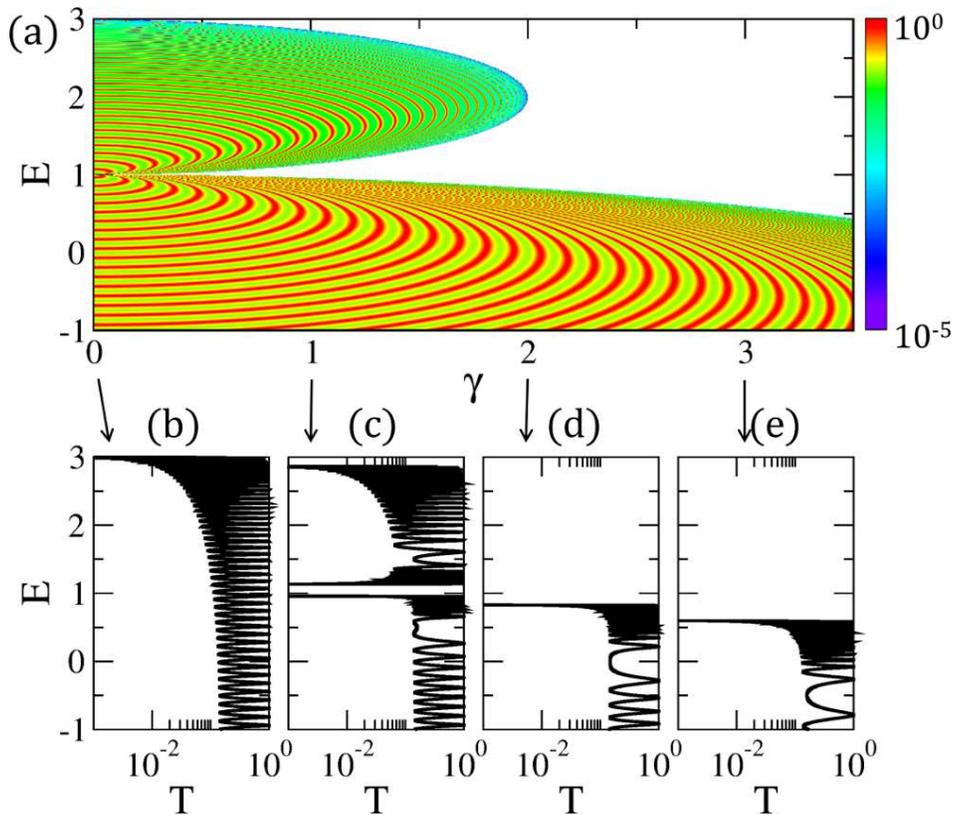}
\caption{(a) Transmission probabilities in the CS lattice with $N=100$ unit cells in ($\gamma$, $E$) space.
Red curves represent high transmission probabilities corresponding to the resonant modes in the finite $\mathcal{PT}$-symmetric CS lattice. 
As $\gamma$ increases, band width decreases and bandgap increases. Ultimately, the upper band disappears when $\gamma > \gamma_c = 2$. 
(b)-(e) Transmission probabilities with different $\gamma$. As $E$ approaches the band edges, the spacings between transmission peaks narrow because of the high density of states. We fix $g=1$ and $V_0 = 10$ throughout this paper.
}
\label{fig2}
\end{figure}

In Fig.~\ref{fig2} (a), the transmission probability $T=\left|t_0\right|^2$ is oscillating as a function of energy $E$, and balanced gain and loss $\gamma$ and the oscillation peaks are indicated resonances with eigenenergies as shown in (b)-–(e). There exists transmission suppression, which makes the boundaries the same as the band edges for the unbroken phase in Fig.~\ref{fig1} (a). Figure~\ref{fig2} (b) shows the transmission probability without imbalanced potentials, which clearly corresponds to the dispersive bands of Fig.~\ref{fig1} (b) because the compact localized states in the flat band are such that they are unavailable for charge transfer. For all cases, the transport phenomena reflects the real eigenenergies of the lattice.
Regarding transmission, it is striking that the broken $\mathcal{PT}$-symmetric phase does not contribute to charge transfer in any energy range, including $E = t$, which is flat band energy when $\gamma = 0$. Consequently, transport is determined by the unbroken phase, having only real energy bands; transport is not determined by the broken phase, even though the complex energy bands are dispersive, as seen in Fig.~\ref{fig1} (c)–-(e). 
It is noteworthy that the behavior of band width depending on balanced gain and loss is totally distinct from precedent bandgap engineerings \cite{Cap87}. The band width narrows as $\gamma$ increases, until all states of the upper band disappear in transmission probability.
After the critical value $\gamma>\gamma_c$, the transmission is suppressed above the remaining lower energy band. The suppressed region is separated by unmeasurable and forbidden regions, which either include resonant states or not, e.g., in the cases of $\varepsilon_{r} < \varepsilon_c$ and $\varepsilon_{r} > \varepsilon_c$, respectively. 

Besides variations in the bandgaps, the resonant levels near the EPs are quite different from those near the band edges. The level spacings are large near the EPs, while the resonant levels are dense near the band edges; this originates from the density of states being inversely proportional to the group velocity, $D(\varepsilon)=D(k)/(d\varepsilon/dk)$. As seen in Fig.~\ref{fig1} (c)–-(e), the group velocity is diverging close to the singular EPs and converging to zero close to the band edges.
Thus, at the EPs, there is no resonant transmission through a lattice with a finite number of sites. In addition, the shifts of resonant energies near the EPs are very sensitive to perturbations.

%Fig3%

\begin{figure}
\centering
\includegraphics[width=0.7\columnwidth]{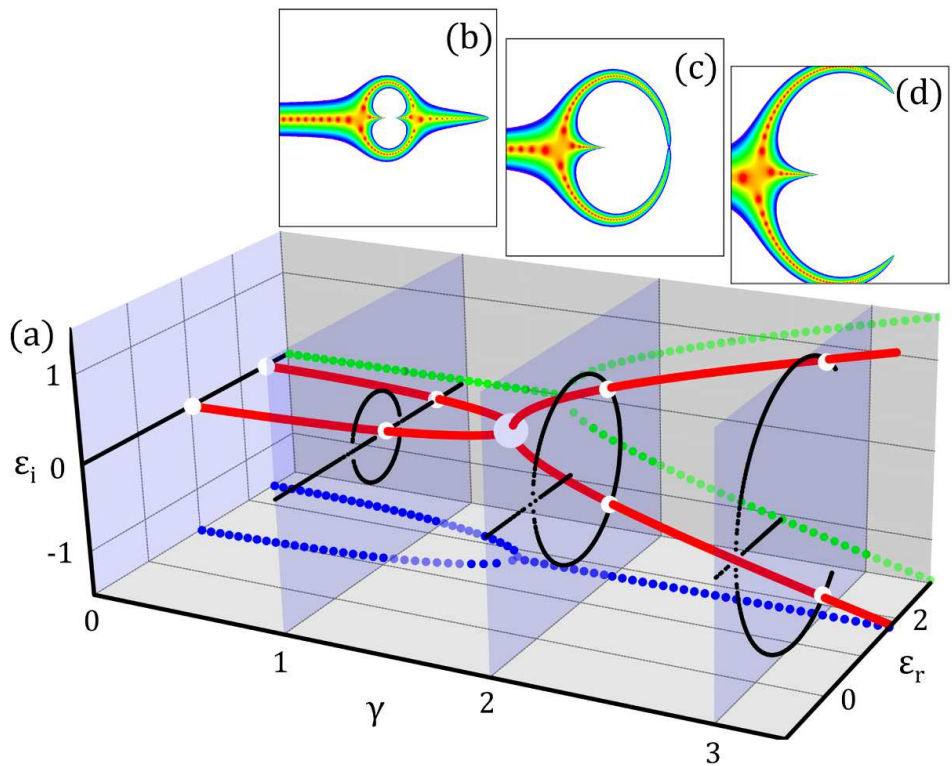}
\caption{(a) Evolution of two selected eigenenergies connected via EP in a CS lattice with $N=100$ unit cells. As $\gamma$ increases, two real eigenenergies (solid red lines) coalesce at an EP (large gray dot), and then split into two complex conjugate eigenenergies from the EP. The two projected figures show the real (blue dotted) and imaginary (green dotted) parts of the complex eigenenergies as a function of $\gamma$. All eigenenergies on the complex planes are shown, which correspond to the complex energy band structures (black dots on the blue panels) in Fig.~\ref{fig1} with corresponding $\gamma$.
(b)-(d) Transmission probabilities on the complex incident energy plane reflecting complex eigenenergies. The red region represents high transmission probability corresponding to the resonant states as in Fig.~\ref{fig2}.
}
\label{fig3}
\end{figure}

{\it Eigenenergy behavior in $\mathcal{PT}$-symmetric cross-stitch lattices} -
Figure~\ref{fig3} (a) shows the evolution of a complex eigenenergy pair of resonant states, connected via an EP, in the quasi-1D lattice with finite unit cells (see Supplementary Information III). As $\gamma$ increases, two real eigenenergies coalesce at an EP before splitting into two complex conjugate eigenenergies. Each pair of eigenenergies is connected via different EPs, of which position is the crossing point between real and complex eigenenergies on the complex energy plane.
The eigenenergies of the finite system overlap the band structures on the complex energy plane (Fig.~\ref{fig3} (a)). The transmission probabilities of the resonant states are also matched on the band structures (Fig.~\ref{fig3} (b)-–(d)) when we {\it control the complex incident energy}, $E=E_r+iE_i$, at a source lead. Control in other words means that we govern energy gain or loss on the lead.

We can see that the transmission probability is suppressed inside the bandgap generated by the unbroken phases. The unmeasurable resonant states, however, are not expelled toward the exterior of the bandgap on real energy but remain inside the bandgap as states containing complex energy. In other words, the resonant states are inflated toward the complex energy plane from flat band energy due to symmetry breaking as $\gamma$ increases.

Finally, let us consider a non-$\mathcal{PT}$-symmetric system with overall gain or loss, $\gamma_{u,d}=-\Gamma\pm\gamma/2$; here, the eigenenergies are shifted in the direction of imaginary energy by the amount of overall gain or loss $\Gamma$ from the eigenenergies of the $\mathcal{PT}$-symmetric system. By changing $\Gamma$ with $\gamma \neq 0$ and $E_i=0$, the transmission probability can be measured as a function of $E_r$ and $\Gamma$, with results consistent with the controlling of complex incident energy in the $\mathcal{PT}$-symmetric system. In other words, the measurement of transport with $\Gamma$ control in the system plays the role of measurement with controlling gain and loss on a lead. The measured states are resonant with eigenenergies of Eq.~(\ref{energy}) as well, $E_r=\varepsilon_{\pm}-i\Gamma$ (see Supplementary Information IV).
Although we lose unitarity by managing $\Gamma$, we can secure significant characteristics of broken $\mathcal{PT}$-symmetric states through the real incident energy.

~~~~~ Compared to typical transport characteristics in Hermitian systems, an intriguing transport phenomenon has been revealed in a non-Hermitian system: the suppression of transmission in an energy range that corresponds to the broken phases between the band edges of the unbroken phases. Such suppression phenomena can be intuitively understood by the concept of group velocity. Because of the imaginary imbalanced potential, the eigenenergies of the non-Hermitian system are given as complex numbers, and accordingly, the group velocity, i.e. $d\varepsilon/dk$, is expected to be complex. While real group velocity corresponds to propagating modes, imaginary group velocity is reflected in evanescent modes. Interestingly, in the broken phase, the group velocity has been found to be a complex number, of which the imaginary part is reflected in the suppression of the transmission due to the attenuation of the evanescent modes.

Quasi-1D lattices with reflection symmetry, such as CS, saw-tooth, or tunable diamond lattices, exhibit flat bands that appear due to destructive interference and local symmetry. These lattices can be easily changed into $\mathcal{PT}$-symmetric lattices by applying gain and loss on the spatially symmetric sites, as in the inset of Fig.~\ref{fig1}. The differences between Hermitian and non-Hermitian imbalanced potentials can be clearly understood by means of detangling the flat bands into Fano lattices \cite{Fla14}. While the Hermitian imbalanced potentials make the coupling between Fano states and continuum real, the $\mathcal{PT}$-symmetric imbalanced potentials produce imaginary coupling between them (see Supplementary Information V).

While $\mathcal{PT}$-symmetry has been studied in a variety of fields, its applications seem to focus on optics, after experimental achievements in $\mathcal{PT}$-symmetric optical waveguides \cite{Guo09, Rue10}, where the control of gain and loss is more widely adopted than in other fields. Recently, it has been reported that the band structure of 2D $\mathcal{PT}$-symmetric photonic crystals can be engineered by controlling gain and loss, as in our results for quasi-1D lattices \cite{Cer16}. Basically, the photonic band structure of $\mathcal{PT}$-symmetric photonic crystals corresponds to the band structure in tight-binding lattice models because of the mathematical analogy between the Helmholtz equation and the Schr{\"o}dinger equation. This analogy also gives the correspondence between the complex refractive index in an optical device and the complex potential in a lattice model.

In summary, a reconfiguration of quantum states in $\mathcal{PT}$-symmetric quasi-one dimensional lattices has been demonstrated, where the quantum states can be controlled by balanced gain and loss. We have explored how the variations of quantum states originate in the transition from the unbroken to broken $\mathcal{PT}$-symmetric phase, via exceptional point. As a result, transmission probabilities in the $\mathcal{PT}$-symmetric system are only determined by bands with an unbroken phase. Conversely, in the case of a non-$\mathcal{PT}$-symmetric system with overall gain and loss, transmission probability is determined by the broken $\mathcal{PT}$-symmetric phase of corresponding $\mathcal{PT}$-symmetric systems with a shift in imaginary energy by the amount of overall gain and loss.

\section*{Acknowledgements}

This work was supported by Project Code (IBS-R024-D1).

\newpage

\section*{SUPPLEMENTARY INFORMATION}

\section{Hamiltonian of a cross-stitch lattice}

The Hamiltonian of the cross-stitch lattice shown in Fig.~1 is given by
\begin{equation}
 E \Psi_j = H_0 \Psi_j + H_1 \Psi_{j+1} + H_1^+ \Psi_{j-1},
 \label{scheq}
\end{equation}
where
\begin{equation}
 H_0 = \left(\begin{array}{cc}
 \epsilon_a & -t \\
 -t & \epsilon_b \\
\end{array}\right),
 H_1 = \left(\begin{array}{cc}
 -d & -d \\
 -d & -d \\
\end{array}\right),
\label{cslattice}
\end{equation}
and $\Psi_j = (\phi_j^a, \phi_j^b)^T$.
We can set $\Psi_{j+1} = \Psi_j e^{ik}$ and $\Psi_{j-1} = \Psi_j e^{-ik}$ due to the translational symmetry of the unit cells. Finally,
\begin{equation}
H = \left(\begin{array}{cc}
 \epsilon_a - 2d \cos{k} & -t -2d \cos{k} \\
 -t -2d \cos{k} & \epsilon_b -2d \cos{k} \\
\end{array}\right).
\end{equation}
Solving the eigenproblem of $H$ when $\epsilon_a = \epsilon_b = 0$ and $t=d=1$, we obtain the band structure for the cross-stitch lattice in Fig.~2 (a) as follows
\begin{equation}
 \varepsilon(k)=-t-4 d \cos{k},   ~\varepsilon_{FB} = t.
\end{equation}

\begin{figure}
\begin{center}
\includegraphics[width=\figsizeone\textwidth]{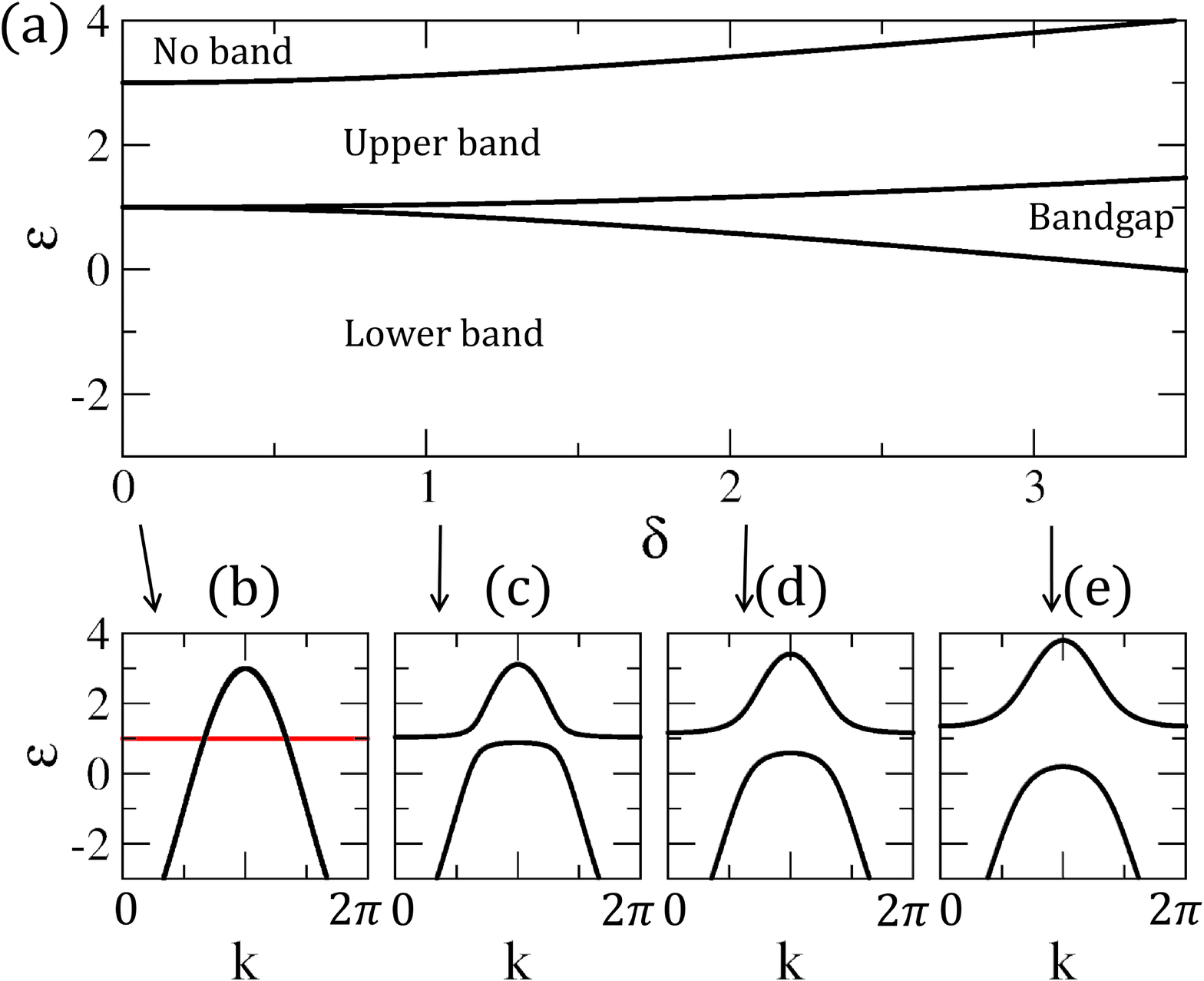}
\caption{
(a) Phase diagram of a cross-stitch lattice with real value perturbations to on-site energies. The real energy on $a$-sites (red dots) and $b$-sites (blue dots) equal $+ \delta /2$ and $- \delta /2$, respectively. Black curves represent the boundaries of the energy bands. There are no electronic states in no-band regions as well as bandgap regions.
(b)-–(d) Energy bands when $\delta$ equals $0$, $1$, $2$, and $3$, respectively.
The red constant energy band in (b) is the flat band.
}
\label{fig_s1}
\end{center}
\end{figure}

We set $\epsilon_a=\delta/2 + i \gamma/2$ and $\epsilon_b=-\delta/2 - i \gamma/2$.
Figure~\ref{fig_s1} (a) shows a phase diagram of a cross-stitch lattice with real value perturbations to on-site energies in ($\delta$, $\varepsilon$) space when $\gamma = 0$. As $\delta$ increases, the size of the energy bandgap increases.

\section{Transmission probability of s-matrix in finite sized cross-stitch lattices}

We now discuss the transport problem in finite sized cross-stitch lattices.
The system has a cross-stitch lattice with $N$ unit cells as shown in Fig.~1, with two leads connected to both $a$- and $b$-sites of the left and right end unit cells of the lattice.
The Hamiltonian of this system is given by
\begin{equation}
 H = H_{cs} + H_{lead} + H_{coupling},
\label{fb_trans2}
\end{equation}
where $H_{cs}$, $H_{lead}$, and $H_{coupling}$ describe the cross-stitch lattice, leads, and coupling between the lattice and leads, respectively.
\begin{eqnarray}
 H_{cs} &=& \sum_{i=1}^{N}H_0 d_{i}^{\dagger} d_{i} + \sum_{i=1}^{N-1} (H_{1} d_{i+1}^{\dagger} d_{i} + h.c.) \\
 H_{lead} &=& -\frac{V_0}{2} \sum_{j \neq 0} (c_{j+1}^{\dagger} c_{j} + h.c.) \\
 H_{coupling} &=& -g ( d_{1}^{\dagger} c_{-1} + c_{1}^{\dagger} d_{N} + h.c.), 
\end{eqnarray}
where $d_{j}^{\dagger}$ ($d_{j}$) and $c_{j}^{\dagger}$ ($c_{j}$) are electron creation (annihilation) operators for the lattice and leads, respectively. $V_0 / 2$ is a hopping strength in the leads and $g$ is a coupling strength between the cross-stitch lattice and leads.
\begin{eqnarray}
 E \phi_{-1} &=& -\frac{V_0}{2} \phi_{-2} - g a_{1} \\
 E a_{1} &=& H_{0} a_{1} + H_{1} a_{2} - g \phi_{-1}\\
 E a_{j} &=& H_{0} a_{j} + H_{1}^{\dagger} a_{j-1} + H_{1} a_{j+1} (2 \leq j \leq N-1) \\
 E a_{N} &=& H_{0} a_{N} + H_{1}^{\dagger} a_{N-1} - g \phi_{1} \\
 E \phi_{1} &=& -\frac{V_0}{2} \phi_{2} - g a_{N}, 
\end{eqnarray}
where
\begin{eqnarray}
 \phi_{j} =& e^{i q j} + r_0 e^{-i q j} & (j < 0) \\
          =& t_0 e^{i q j} & (j > 0). 
\end{eqnarray}
$\left|r_0\right|^2$ and $\left|t_0\right|^2$ are reflection and transmission probabilities, respectively, and $\left|r_0\right|^2 + \left|t_0\right|^2 = 1$ in Hermitian cases.

Finally, we obtain the equations as follows
\begin{eqnarray}
 -\frac{V_0}{2} &=& \frac{V_0}{2} r_0 - g a_1 \\
 g e^{-i q} &=& - g e^{i q} r_0 + (H_{0} - E) a_{1} + H_{1} a_{2} \\
 0 &=& H_{1}^{\dagger} a_{j-1} + (H_{0} - E) a_{j} + H_{1} a_{j+1} \\
 0 &=& H_{1}^{\dagger} a_{N-1} + (H_{0} - E) a_{N} - g e^{i q} t_0 \\
 0 &=& \frac{V_0}{2} t_0 - g a_{N}, 
\end{eqnarray}
where 
\begin{eqnarray}
 e^{\pm i q} = - \frac{E}{V_{0}} \pm i \sqrt{1 - \left|\frac{E}{V_{0}}\right|^2}. 
\end{eqnarray}
Finally, we can obtain $R$ and $T$ from the following equation
\begin{equation}
 \left(\begin{array}{c}
 -\frac{V_0}{2} \\
 G_{2-} \\
 0 \\
 \vdots \\
 0 \\
 0 \\
 0 \\
\end{array}\right)=
 \left(\begin{array}{ccccccc}
 \frac{V_0}{2} & G_1 & & & & & \\
 G_{2+} & H_0 - E I & H_1 & & & &  \\
  & H_1^{\dagger} & H_0 - E I & H_1 & & &  \\
  & & \ddots & \ddots & \ddots && \\
  &&& H_1^{\dagger} & H_0 - E I & H_1 & \\
  &&&& H_1^{\dagger} & H_0 - E I & G_{2+} \\
  &&&&& G_1 & \frac{V_0}{2} \\
\end{array}\right)
\left(\begin{array}{c}
 r_0 \\
 a_1 \\
 a_2 \\
 \vdots \\
 a_{N-1} \\
 a_{N} \\
 t_0 \\
\end{array}\right),
\end{equation}
where
\begin{equation}
G_1 = - g
\left(\begin{array}{cc}
 1 &  1 \\
\end{array}\right),
G_{2\pm} = \mp g e^{\pm iq}
 \left(\begin{array}{c}
 1 \\
 1 \\
\end{array}\right),
\end{equation}
and $H_0$ and $H_1$ are $2 \times 2$ matrices in the case of cross-stitch lattice with two leads coupled to both $a$- and $b$-sites of the end unit cells.

\begin{figure}
\begin{center}
\includegraphics[width=\figsizeone\textwidth]{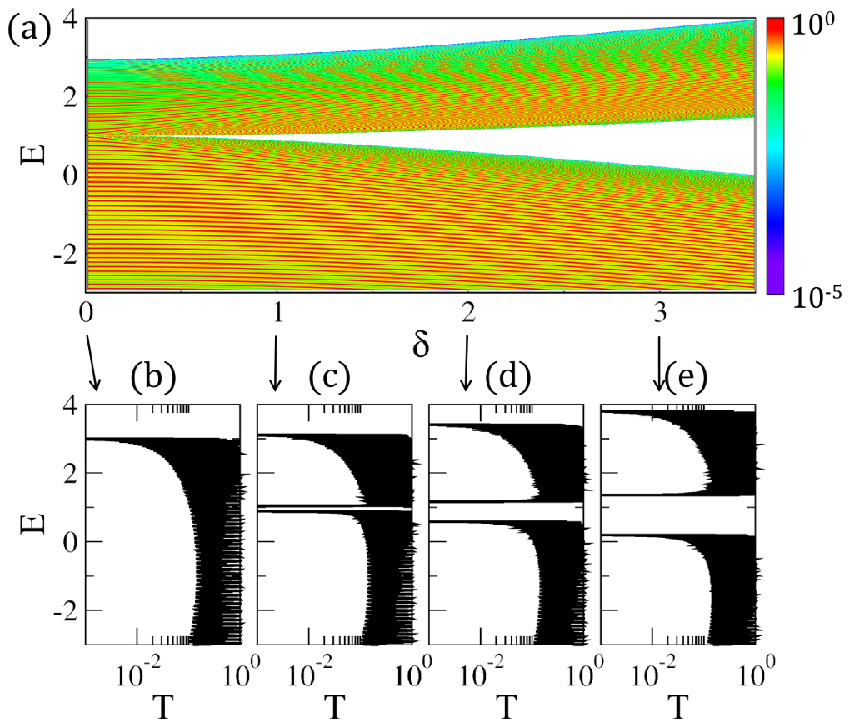}
\caption{
(a) Transmission in ($\delta$, $E$) space when $\gamma = 0$.
Red curves represent high transmissions corresponding to the resonant modes in finite sized cross-stitch lattices. The transmission is smaller than $10^{-5}$ in the white region, which is a bandgap in transmission.
The resonant states near the boundaries of the bands are not clear because of the high density of states.
As $\delta$ increases, the sizes of the bands are mostly maintained, and the energy states related to the resonant modes do not disappear.
(b–-e) Transmissions when $\delta$ equals $0$, $1$, $2$, and $3$, respectively. As $E$ approaches the boundaries of the bands, the spacings between transmission peaks decrease because of the high density of states.
}
\label{fig_s2}
\end{center}
\end{figure}

Figure~\ref{fig_s2} (a) shows the transmission in ($\delta$, $E$) space when $\gamma = 0$. In the Hermitian case, resonant modes inside the bands do not disappear, so the energy bands almost maintain their widths and density of states, irrespective of perturbation strength. Figure~\ref{fig_s2} (b), (c), (d) and (e) are transmissions when $\delta$ equals $0$, $1$, $2$, and $3$, respectively.

\section{Eigenenergies in finite sized cross-stitch lattices}

We now obtain eigenenergies in finite sized cross-stitch lattices.
From Eq.~(\ref{scheq}) and Eq.~(\ref{cslattice}), $H$ for the cross-stitch lattice with $N$ unit cells is given by
\begin{equation}
 H = \left(\begin{array}{ccccc}
 \ddots & & & & \\
  & H_0 & H_1 & &  \\
  & H_1^+ & H_0 & H_1 &   \\
  & & H_1^+ & H_0 &  \\
  & & & & \ddots   \\
\end{array}\right).
\end{equation}
Solving this $2N \times 2N$ matrix, we can obtain $2N$ eigenenergies.
For instance, in the case of $\epsilon_a=\epsilon_b$, each band has $N$ corresponding eigenenergies, i.e., $N$ eigenenergies correspond to the flat band and $N$ eigenenergies relate to the dispersion band.

\section{Transmission in a non-$\mathcal PT$-symmetric non-Hermitian system}

\begin{figure}
\begin{center}
\includegraphics[width=0.5\columnwidth]{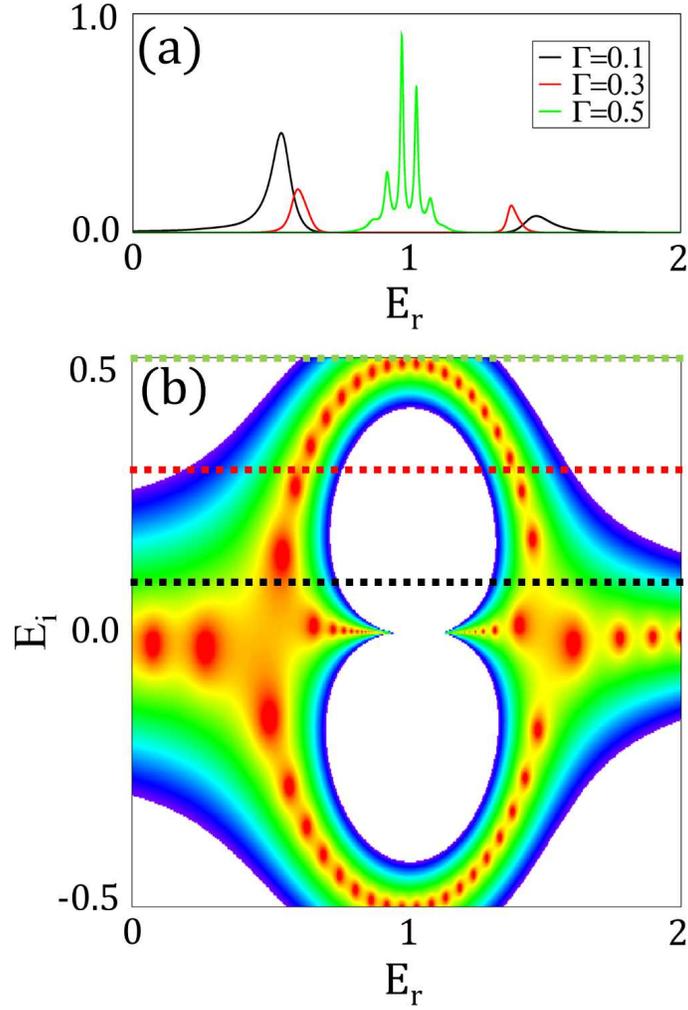}
\caption{
(a) Transmission probabilities when $\Gamma=0.1$ (black), $0.3$ (red), and $0.5$ (green).
(b) Transmission probability on the complex incident energy plane when $\gamma = 1$. The red regions represent high transmission probability corresponding to the resonant states.
The black, red, and green dotted lines represent $E_{i}=0.1$, $0.3$, and $0.5$, respectively.
}
\label{fig_s3}
\end{center}
\end{figure}

Let us consider the transmission probability in a cross-stitch lattice with $\gamma = 1$ and different $\Gamma$. When $\Gamma=0.1$, there are two peaks at $E_{r} \sim 0.54$ and $E_{r} \sim 1.47$ in transmission with real incident energy, which correspond to the transmission probability on the line of $E_{i} = 0.1$ when $\gamma = 1$ in Fig.~\ref{fig_s3} (b). When $\Gamma=0.3$ and $0.5$, there are also peaks in transmission with real incident energy, which correspond to the transmission probability on the lines of $E_{i} = 0.3$ and $0.5$, respectively.

\section{Detangling cross-stitch lattices into Fano lattices}

\begin{figure}
\begin{center}
\includegraphics[width=\figsizeone\textwidth]{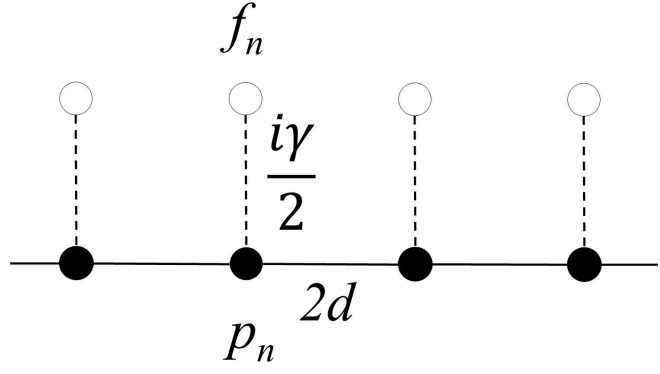}
\caption{
Fano lattices detangled from a $\mathcal PT$-symmetric cross-stitch lattice.
}
\label{fig_s4}
\end{center}
\end{figure}

Following Ref.~\cite{Fla14}, we can detangle $\mathcal PT$-symmetric cross-stitch lattices into Fano lattices. The amplitude equations for the Hamiltonian of the cross-stitch lattice (Eq.~(\ref{scheq})) are
\begin{eqnarray}
E a_n &=& \epsilon_n^a - d a_{n+1} - d a_{n-1} - d b_{n+1} - d b_{n-1} - t b_{n}, \\
E b_n &=& \epsilon_n^b - d b_{n+1} - d b_{n-1} - d b_{n+1} - d b_{n-1} - t a_{n}.
\end{eqnarray}
If $\epsilon_a = \epsilon_b$, there is one flat and one dispersion band. From these equations, we obtain a 1D lattice with hopping strength $2 d$, $p_n$, and side-coupled Fano states $f_n$,
\begin{eqnarray}
E p_n &=& (\epsilon_n^{+} - t) p_n +\epsilon_n^{-1} f_n - 2 d (p_{n+1} + p_{n-1}), \\
E f_n &=& (\epsilon_n^{+} + t) f_n + \epsilon_n^{-1} p_n,
\end{eqnarray}
where
\begin{eqnarray}
p_n &=& \frac{1}{\sqrt{2}} (a_n + b_n), ~~~~~ f_n = \frac{1}{\sqrt{2}} (a_n - b_n), \\
\epsilon_n^{+} &=& \frac{1}{2} (\epsilon_n^a + \epsilon_n^b), ~~~~~  \epsilon_n^{-} = \frac{1}{2} (\epsilon_n^a - \epsilon_n^b).
\end{eqnarray}
Figure~\ref{fig_s4} shows the Fano lattices detangled from a cross-stitch lattice. Horizontal couplings are $2 d$ and vertical coupling $\epsilon_n^{-}$. In the case of Hermitian perturbation, $\delta \neq 0$ and $\gamma = 0$, the real energy bands originate from the Hermitian (real value) coupling $\epsilon_n^{-} = \delta/2$ between the Fano states and 1D lattice. In the case of non-Hermitian perturbation, $\delta = 0$ and $\gamma \neq 0$, however, the complex energy bands originate from the non-Hermitian (imaginary value) coupling $\epsilon_n^{-} = i \gamma/2$ between the Fano states and 1D lattice. As a result, the difference between Hermitian and non-Hermitian perturbations depends on whether the couplings between Fano states and 1D lattice in the Fano lattices detangled from a cross-stitch lattice are real or imaginary.

\end{document}